\documentclass[amsfonts,amssymb,epsfig,superscriptaddress,showpacs,doublecol]{epl2}

\usepackage{graphics,graphicx}
\usepackage{amsmath,bbm}
\usepackage{amstext}
\usepackage{amssymb}

\begin{document}

\title{Quantum Game of Life}
\author{D.~Bleh \and T.~Calarco \and S.~Montangero}
\institute{Institut f\"ur Quanteninformationsverarbeitung
Albert-Einstein-Allee 11
D-89069 Ulm, Germany.}

\abstract{
We introduce a quantum version of the Game of Life and we use
it to study 
the emergence of complexity in a quantum world. We show 
that the quantum evolution displays signatures of complex behaviour 
similar to the classical one, however a regime exists, where the
quantum Game of Life creates more complexity, 
in terms of diversity, with respect to the {corresponding classical 
reversible one}.}

\maketitle

The Game of Life (GoL) has been proposed by Conway in 1970 as
a wonderful mathematical game which can describe the appearance of
complexity and the 
{evolution of ``life''} under some simple rules~\cite{conway}. 
Since its introduction it has attracted a lot of attention{, 
as despite its simplicity,} it can reveal complex patterns with
unpredictable evolution: From the very beginning a lot of structures
have been identified, from simple
blinking patterns to complex evolving figures which have been named
``blinkers'', {``gliders''} up to ``spaceships'' due to their 
appearance and/or dynamics~\cite{figs}.
{The classical GoL has been {the} subject of {many}
studies: It has been shown that cellular automata
defined by the GoL {have}  the power of a Universal Turing
machine, that is, anything that can be computed algorithmically can be
computed within Conway's GoL~\cite{Adamatzky,turing}. 
Statistical analysis and analytical descriptions of the GoL 
have been performed; many generalisations or modifications 
of the initial game have been
introduced as, for example, a simplified one dimensional 
version of the GoL and a semi-quantum 
version~\cite{analitic,oned,SQgol}.}
Finally, to allow a statistical mechanics description of the GoL, 
stochastic components have been added~\cite{SMgol}.

In this letter, we bridge the field of {complex systems}
 with quantum mechanics 
introducing a purely quantum GoL and we investigate its dynamical 
properties. We show that it displays interesting
features in common with its classical counterpart, in particular
regarding the variety of supported dynamics and different behaviour. 
The system converges to a quasi-stationary configuration 
in terms of macroscopic variables, and these stable configurations
depend on the initial state, e.g. the initial density of ``alive''
sites for random initial configurations. We show that 
simple, local rules support complex behaviour and that the 
diversity of the structures formed in the steady state 
resembles that of the classical GoL, however a regime exists where 
quantum dynamics {allows more diversity to be created} than possibly 
reached by the classical one. 

\begin{figure}
  \begin{center}    
    \includegraphics[width=5cm]{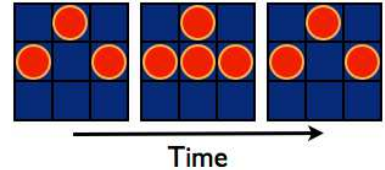}
    \caption{Example of the evolution of the GoL described by
      Hamiltonian~(\ref{ham}) for a simple initial configuration.
      Empty (blue) squares are ``dead'' sites, coloured (red) 
      ones are ``alive''.}
    \label{rule}
  \end{center}
\end{figure}
The universe of the original GoL is an infinite two-dimensional 
orthogonal grid of square cells with coordination number eight, 
each of them in one of two possible states, alive or dead~\cite{conway}. 
At each step in time, the pattern present 
on the grid evolves instantaneously following {simple} rules:
any dead cell with exactly three live neighbours comes to
life; any live cell with less than two or more than three live 
neighbours dies as if by loneliness or overcrowding.
{As already pointed out in~\cite{SQgol}, the rules of the GoL are
irreversible, thus their generalisation to the quantum case   
implies rephrasing them to make them compatible with a quantum
reversible evolution}. The system under study is a collection of
two-level quantum systems, with two possible orthogonal states,
namely the state ``dead'' ($|0\rangle$) and ``alive'' ($|1\rangle$). 
Clearly, differently from the classical case, a site can be also in a
{superposition} of the two possible classical states. 
\begin{figure*}[t]
  \begin{center}    
 \includegraphics[width=18cm]{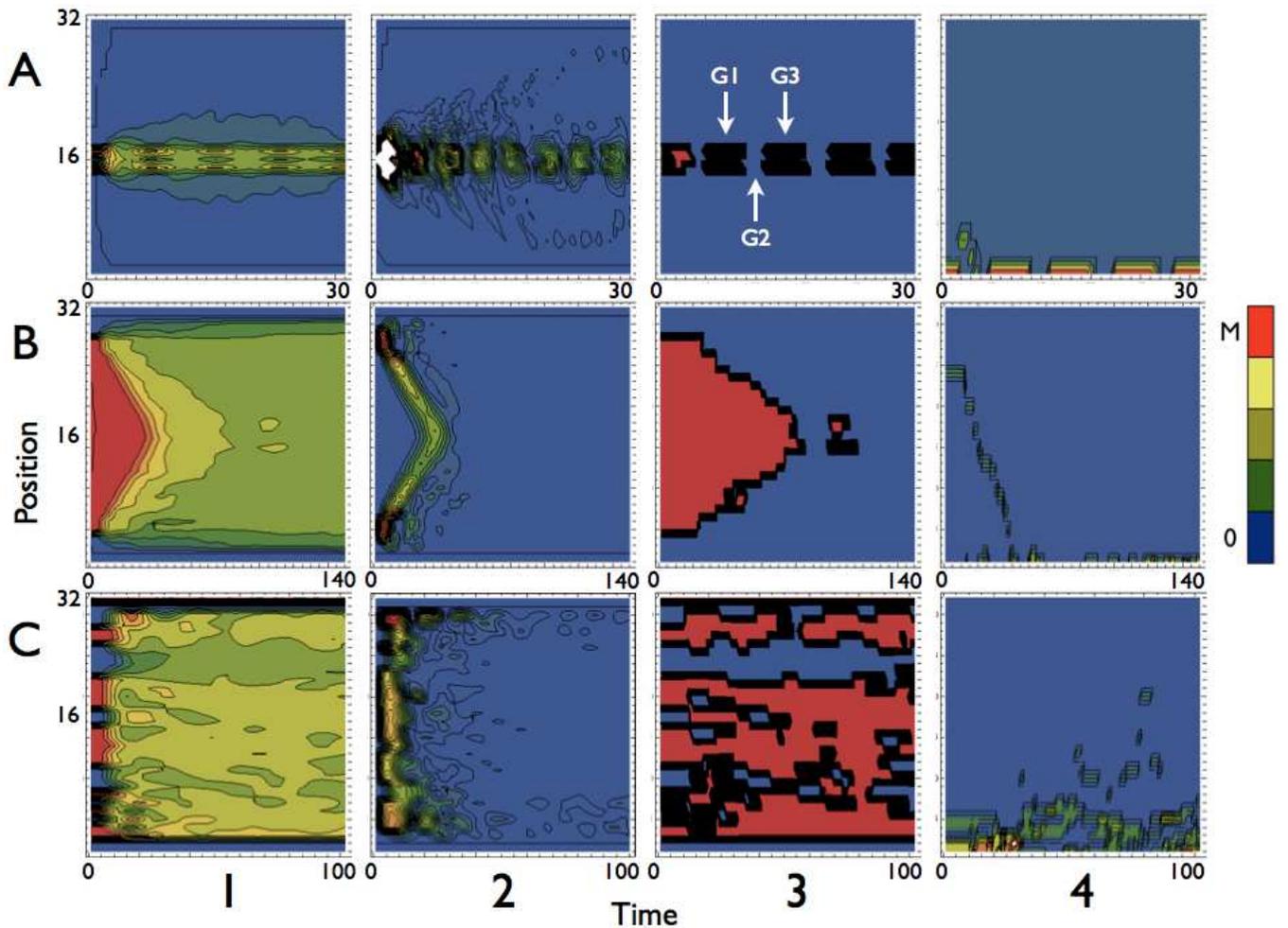}
    \caption{Colour on-line. From left to right: Countour plot of the
      time evolution of the populations $\langle n_i(t) \rangle$
      (column 1), 
      visibility $v_i(t)$ (column 2), 
      discretized populations $\mathcal{D}_i(t)$ (column 3)
      and clustering $\mathcal{C}(\ell,t)$ (column 4)
      for three different initial configurations: four
      alive sites separated by two dead ones (A), twenty-four alive sites
      grouped together (B) and a random initial configuration (C). 
      {Time is reported on the x-axis (in arbitrary units), and
        position (cluster size) $i= 1, \dots ,L$  
      on the y-axis in columns one to three (four).
      Arrows in panel 3A highlight the three subsequent generations
      of a ``blinker'' reported schematically in Fig.~\ref{blink}.}
      The colour code goes from zero 
      to $M=1$ ($M=4$ for the clustering and to $M=.1$ for the
      visibility), from blue through green to red.}
    \label{timeev}
  \end{center}
\end{figure*}
The dynamics is defined as follows in terms of the GoL language:
a site with two or three neighbouring alive sites is active, 
where active means that it will come to life and eventually 
die on a typical timescale $T$ (setting the problem
timescale, or time between subsequent generations). 
That is, if maintained active by the surrounding conditions, 
the site will complete a full rotation, if not, it is 
{``frozen''} in its
state. Stretching the analogy with Conway's GoL to the limit, 
we are describing the evolution of a Virus culture: each individual
undergoes its life cycle if the environment 
allows it, {otherwise it hibernates in its current state and
waits for conditions to change such that the site may become active again.} 
This slight modification {allows us} to recover the reversibility of
the dynamics and to introduce a quantum model that, as we shall see,
reproduces most of the interesting complex behaviour of the classical
GoL from the point of view of a classical observer. 
However, its evolution is purely 
quantum and thus we are introducing a tool that will allow to 
study the emergence of complexity from the 
quantum world. 

\section{Model}  
The Hamiltonian describing the aforementioned model is given by
\begin{equation}
H = \sum_{i=3}^{L-2}(b_i + b_i^\dagger ) \cdot \left(\mathcal{N}^3_i+
\mathcal{N}^2_i \right)
\label{ham}
\end{equation}
where {$L$ is the number of sites;} 
$b$ and $b^\dagger$ are the usual annihilation and creation
operators ($\hbar=1$); the operators 
$\mathcal{N}^2_i = \sum_P n_{\alpha}n_{\beta}\bar n_{\gamma} \bar n_{\delta}$ and 
$\mathcal{N}^3_i=\sum_{P'} n_{\alpha} n_{\beta}  n_{\gamma} \bar n_{\delta}$ 
($n=b^\dagger b$, $\bar n = 1-n$, the indices $\alpha, \beta, \gamma,
\delta$ label the four neighbouring sites) 
count the population present in the four neighbouring sites (the sum
runs on every possible permutation $P$ and $P'$ of the positions of
the $n$ and $\bar n$ operators) and $\mathcal{N}^2$ ($\mathcal{N}^3$)
gives the null operator if the population is different
from two (three), the identity otherwise. For classical states, as for
example an initial random configuration of dead and alive states, 
the Hamiltonian~(\ref{ham}) is, at time zero, $H_{Active}= b_i + b_i^\dagger $
on the sites with two or three alive neighbours 
and $H_{Hibernate}= 0$ otherwise. If the Hamiltonian would remain
constant, every active site would oscillate forever while the
hibernated ones would stand still. On the contrary 
as soon as the evolution starts, the state evolves into a  
superposition of possible classical configurations, resulting in a 
complex dynamics as shown below and the interaction between sites
starts to play a role. 
{Thus, the Hamiltonian introduced in Eq.~(\ref{ham}) induces 
a quantum dynamics that resembles the rules of the GoL: a site with less
than two or more than three alive neighbouring sites "freezes'' while, on
the contrary, it ``lives''. The difference with the classical game --
connected to the reversibility of quantum dynamics -- 
is that ``living'' means oscillating with a typical timescale between
two possible classical states (see e.g. Fig.~\ref{rule}.}

\begin{figure}
  \begin{center}    
    \includegraphics[width=4cm]{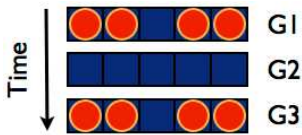}
    \caption{Schematic representation of a one-dimensional
      time-evolution of the discretized population 
      $\mathcal{D}_i(t)$ of a 
      ``blinker'' (case A of Fig.~\ref{timeev}). 
      From left to right the states of subsequent generations are
      sketched.  
      Empty (blue) squares are ``dead'' sites, coloured (red) 
      ones are ``alive''}
    \label{blink}
  \end{center}
\end{figure}
\section{Dynamics} To study the quantum GoL dynamics we employ the time
dependent Density Matrix Renormalization group (DMRG). Originally
developed to investigate condensed matter
systems, the DMRG and its time dependent extension have been proven to
be a very powerful method to numerically investigate many-body quantum
systems~\cite{white,daley,revDMRG, dechiara}. As it is possible to use it efficiently only in
one-dimensional systems, we concentrate to the one-dimensional version
of the Hamiltonian~(\ref{ham}): the operators  $\mathcal{N}^2$
and $\mathcal{N}^3$ count the populated sites on the nearest-neighbour 
and next-nearest-neighbour sites and thus $\alpha=i-2, \beta=i-1, \gamma=i+1,
\delta=i+2$. Note that it has been shown that the main
statistical properties of the classical GoL are the same in 
both two- and one-dimensional versions~\cite{oned}.

{To describe the system dynamics we introduce different quantities
that characterise in some detail the system evolution.}
We first concentrate on the population dynamics, measuring the
expectation values of the number operator at every site $\langle
n_i(t) \rangle$. This clearly gives a picture of the ``alive'' 
and ``dead'' sites as a function of time, as it gives the probability 
of finding a site in a given state when measured. That is, if we observe 
the system at some final time $T_f$ we will find dead or alive sites
according to these probabilities. In Fig.~\ref{timeev} we show three
typical evolutions (leftmost pictures): configuration $A$
corresponds to a ``blinker'' where two couples of nearest-neighbour
sites oscillate regularly between dead and alive states {(a schematic
representation of the resulting dynamics of the discretised population 
 $\mathcal{D}_i(t)$ is reproduced also in Fig.~\ref{blink})}; 
configuration $B$ is
a typical overcrowded scenario where twenty-four ``alive'' sites
disappear leaving behind only some residual activity; finally a
typical initial random configuration ($C$) is shown.
Notice that in all configurations it is possible to identify the
behaviour of the wave function tails that propagate and generate
interference effects. These effects can be highlighted by computing
the visibility of the dynamics, the maximum 
variation of the populations within subsequent generations, 
defined as:
\begin{equation}
v_i(t)= |\max_{t'} n_i(t') - \min_{t'} n_i(t')|;
t'\in[t-\frac{T}{2};t+\frac{T}{2}]; 
\end{equation}
{that is, the visibility at time $t$ reports the maximum variation of the
population in the time interval of length $T$ centered around $t$.} 
The visibility clearly follows the preceding dynamics (see
Fig.~\ref{timeev}, second column) and identifies the presence 
of ``activity'' in every site.

\begin{figure}
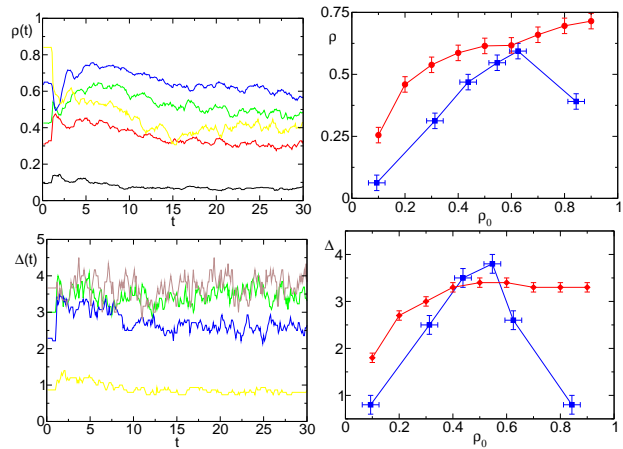

  \begin{center}    
     \includegraphics[width=4cm]{./avpop.eps}
     \includegraphics[width=4cm]{./avpopinf.eps}
\\
     \includegraphics[width=4cm]{./avdiv.eps}
     \includegraphics[width=3.9cm]{./avdivinf.eps}
    \caption{Left: Average population $\rho(t)$ (upper)
      and diversity $\Delta(t)$ (lower) as a function of time {for
      different initial population density $\rho_0$}. 
    Right: Equilibrium average population $\rho$ (upper) and diversity
    $\Delta$ (lower) for the quantum (blue squares) and classical (red
    circles) GoL {as a function of the initial population density
    $\rho_0$}. Simulations are performed with a t-DMRG at third
    order, Trotter step $\delta t=10^{-2}$, truncation dimension
    $m=30$, size $L=32$,  averaged over up to thirty 
    different initial configurations.}
    \label{stat}
  \end{center}
\end{figure}

To stress the connections and comparisons with the original GoL we
introduce a classical figure of merit (shown in the third column of
Fig.~\ref{timeev}):  we report a
discretized version of the populations as a function of time
($\mathcal{D}_i(t)=1$ for $n_i(t) > 0.5$ and $\mathcal{D}_i(t)=0$
otherwise). {Notice that $\mathcal{D}_i(t)$ gives the most
probable configuration of the system after a measurement
on every site in the basis $\{|0\rangle, |1\rangle\}$.} 
{Thus, we recover} a ``classical'' view of the
quantum GoL with the usual definition of site status. For example, 
configuration $A$ is a ``blinker'' that changes status at every
generation {(see Fig.\ref{timeev} and \ref{blink})}. 
More complex configurations appear in the other two
cases. The introduction of the discretized populations $\mathcal{D}_i$
can also be viewed as a new definition of ``alive'' and ``dead'' 
sites from which we could have started from the very beginning to
introduce a stochastic component as done in~\cite{SMgol}. 
This quantity allows analysis to be performed as usually done on
the classical GoL and to stress the similarities between the quantum and
the classical GoL. 
Following the
literature to quantify such complexity, we compute the clustering
function
$\mathcal{C}(\ell,t)$ that gives the number of clusters of 
neighbouring ``alive'' sites of size $\ell$ as a function of time~\cite{oned}. 
{For example, the function $\mathcal{C}(\ell,t)$ for a uniform
distribution of ``alive'' sites would be simply $\mathcal{C}(L)=1$ and
zero otherwise while a random pattern would result in a random cluster
function.} This function characterises the complexity of the evolving patterns,
e.g. it is oscillating between zero- and two-size clusters for the
initial condition $A$, while it is much more complex for the random
configuration $C$ (see Fig.~\ref{timeev}, rightmost column). 
 
\section{Statistics} 
To characterise the statistical properties of the quantum GoL we study 
the time evolution of different initial random configurations as a 
function of the initial density of alive sites. {We concentrate 
on two macroscopic quantities: the density of the sites 
that if measured would with higher probability result in ``alive'' states
\begin{equation}
\rho(t)=\sum_i \mathcal{D}_i(t)/L;
\end{equation}
and the diversity 
\begin{equation}
\Delta(t)= \sum_\ell\mathcal{C}(\ell,t), 
\end{equation}
the number of different cluster sizes that are present 
in the systems, that quantifies the complexity of the 
generated dynamics~\cite{oned, SMgol}.} Typical results,
averaged over different initial configurations, are shown in
Fig.~\ref{stat} (left). As it can be clearly seen 
the system equilibrates and the density of states as well as the
diversity reach a steady value. This resembles the typical
behaviour of the classical GoL where any typical initial random
configuration eventually equilibrates to a stable configuration. 
Moreover, we compare the quantum GoL with a classical {reversible}
version of GoL corresponding to that introduced here: 
at every step a cell changes its status if and only if 
within the first four neighbouring cells only three or two are alive. 
Notice that, the evolution being unitary and thus reversible, 
the equilibrium state locally changes with time, however the 
macroscopic quantities reach their equilibrium values {that 
depend} non trivially only on the initial population density. 
{In fact, for the classical game, we 
were able to check that the final population density is independent
of the 
system size while the final diversity scales as $L^{1/2}$ (up
to $2^{10}$ sites, data not shown).} 
Moreover, the {time needed to reach equilibrium}
 is almost independent of the system
size and initial population density. 
{These results on the scaling of classical system properties
support the conjecture that our findings for the quantum case 
will hold in general}, 
while performing the analysis for bigger system sizes is highly demanding. 
A detailed analysis of the size scaling of the system properties
will be presented elsewhere. 
In Fig.~\ref{stat} we report the final (equilibrium) 
population density (right upper) and diversity (right lower) 
as a function of the initial population density for both the classical
and the quantum GoL for 
systems of $L=32$ cells. The equilibrium population density $\rho$
is a non linear function of the initial one $\rho_0$ in both cases:
the classical one has an initial linear dependence up to half-filling
where a plateau is present up to the final convergence to unit filling
for $\rho_0=1$. Indeed, the all-populated configuration is 
a stable system configuration. The quantum GoL follows a 
similar behaviour, with a more complex 
pattern. Notice that here a first signature of quantum behaviour is
present: the steady population density reached by the quantum
GoL is always smaller than its classical counterpart. This is
probably due to the fact that the evolution is not completely captured
by this classical quantity: the sites with population below half
filling, i.e. the tails of the wave functions, are described as
unpopulated by $\mathcal{D}_i$. However, this missing population 
plays a role in the evolution: within the overall superposition of
basis states, a part of the probability density (corresponding to the
states where the sites are populated)
undergoes a different evolution than the classical one. 
In general, the quantum system is effectively more populated than 
the classical $\rho$ indicates. This difference in the
quantum and classical dynamics is even more evident in the dependence of the
equilibrium diversity on the initial population density
$\rho_0$. In the classical case the maximum diversity is slightly
above three: on average, in the
steady state, there are no more than about three different cluster
sizes present in the system independently of the initial
configuration. On the contrary --in the quantum case-- the
maximal diversity is about four, increasing the information content 
(the complexity) generated by the evolution by about $10-20\%$. 

{These findings are a signature of the difference between quantum and
classical GoL. In particular} we have shown that the quantum GoL has a
higher capacity of generating diversity than the {corresponding}
classical one. This property arises from the possibility of having quantum
superpositions of states of single sites. Whether purely quantum
correlations (entanglement) play a crucial role is under
investigation. Similarly, as there is some arbitrariness 
in our definition of the quantum GoL, the investigation of 
possible variations is left for future work.

The investigation presented here fits perfectly as
a subject of study for quantum simulators, like for example cold atoms
in optical lattices. Indeed, the five-body Hamiltonian~(\ref{ham}) 
can be written in {pseudo spin-one-half operators (Pauli matrices)} and 
thus it can be simulated along the lines
presented in~\cite{cirac}. In particular, these simulations would give
access to investigations in two and three dimensions that are
not feasible by means of t-DMRG~\cite{revDMRG}. 

In conclusion we note that this is {one of the few
available simulations of a many-body quantum game scalable in the number 
of sites~\cite{qgames,mbqg,chen04}}. With a straightforward generalisation 
(adding more than one possible strategy defined in
Eq.~(\ref{ham})) one could study also different
many-player quantum games. This approach will allow 
different issues {to be studied} related 
to many-player quantum games such as the
appearance of new equilibria and their thermodynamical
properties. 
Moreover, the approach introduced here shows that one might 
investigate many different aspects of many-body quantum systems with
the tools developed in the field of complexity and dynamical systems: 
In particular,
the relations with Hamiltonian quantum cellular automata 
in one dimension and quantum games~\cite{qgames,nagaj}.
Finally, the search for the possible existence of self-organised
criticality in these systems along the lines of similar investigations 
in the classical GoL~\cite{SOC}, if successful, would be the 
first manifestation of such effect in a quantum system and might 
have intriguing implications in quantum gravity~\cite{smolin,gupta}.

After completing this work we became aware of another work on the same
subject~\cite{arXive}.

We { acknowledge} 
interesting discussions and support by R.~Fazio and M.B.~Plenio,
the SFB-TRR21, the EU-funded projects 
AQUTE, PICC for funding, the BW-Grid for computational
resources, and the PwP project for the t-DMRG code (www.dmrg.it).

\end{document}